# Left-Handed Representation in Top 100 Male Professional Tennis Players: Multi-Disciplinary Perspectives

Boris Bačić and Ali H. Gazala

School of Engineering, Computer and Mathematical Sciences
Auckland University of Technology
Auckland, New Zealand
{`boris.bacic, ahaiderh`}`@aut.ac.nz`

**Abstract**. A commonly held opinion is that left-handed tennis players are overrepresented compared to the percentage of left-handers within the general population. This study provides the domain insights supported by data analysis that could help inform the decision of parents and coaches considering whether a child should start playing tennis as left- or right-handed when there is no strong arm-handed dominance. Compared to the commonly cited figure of about 10% of left-handed male population, data analysis from the official ATP web site for the top 100 ranked tennis players over the past decades (1985-2016) shows evidence of overrepresentation of left-handed elite tennis players (about 15%). The insights and data analysis can inform the handedness decision, advance coaching and strategic game concepts, enhance media coverage/analytics, left-handed facts and statistics, and inform tennis equipment manufacturing.

**Keywords:** Data analytics, text mining, sport science, sport coaching, sport equipment manufacturing, media coverage.

## 1 Introduction

At present, research [1-3] suggests that about 10% of the population is left-handed. However, there are no definitive answers regarding the influence of society or religion in diverse cultures and the degree of ambidexterity of global population in sports such as tennis. For a person who is considering to start playing tennis or for parents observing their child handedness preferences in addition to family history there are also simple tests e.g. [4] to estimate *laterality index* (LI) as a measure of cerebral asymmetry that may indicate how hard it may be for an individual to adapt to playing sports as left-handed. With a growing number of official web resources providing public sports data records and libraries in various programming languages that enable information extraction or automation of data retrieval from websites, it is possible to summarise and extract knowledge from data to answer questions related to handedness in sports.

## 1.1 Data Acquisition Methods in Text Mining

In the age of big data, researchers and entrepreneurs can take advantage of a myriad of publicly accessible data sources. The sharp increase in data collection and integration tools during the past decade has created many opportunities to derive insights from integrated datasets. The emerging computing discipline of text mining has its roots in data analytics, big data, on-line data analysis, and data base and web design. Text mining unifies three distinct areas: (1) data acquisition; (2) quantifying text/feature extraction techniques and (3) analysis and model design. Data analysis and model design includes both off-line and on-line/real time data interpretation. Examples of data acquisition include web scraping and integration of web sites with data mining tools and libraries via application programming interfaces API e.g. Facebook. For cases in which one-off web site table analysis is needed, Microsoft's Excel provides simple HTML table data into spreadsheet conversion; import/export to CSV text format; internal representation of spreadsheet to data table (enabling rudimentary column analysis such as producing a list of distinct values for each column); and built-in functions and macro writing capability for expressing conversions such as text to categories or 'name=value[unit]' for further numerical analysis. For on- and off-line data analysis using API or web scraping technique, a data analyst must familiarise him/herself with published API libraries or the design elements of the web sites that can provide the required data. Another perhaps more technically challenging, time consuming and generally less popular alternative is to program custom-built software that could transfer data from screen into text from various programmes and reports (e.g. from image screenshots or pdf reports). Such opportunities for text mining could be found in the growing domain of augmented sport coaching technology/ubiquitous systems, including wearable or attached devices to sport equipment [5]. By using text mining approaches to obtain data in augmented sport coaching contexts, it is possible to extend the functionality of sport equipment that would otherwise restrict users from owning and processing their own motion data collections locally rather than on a third-party cloud, hence enabling research in sport science and rehabilitation.

In this study, it is contended that a web scraping technique is a universal scientific tool for data import from web sites emulating user interaction for data retrieval thus enabling multi-disciplinary research. The expected benefits from such automated web data retrieval include potential improvements in terms of reducing human error of manual data copying and recording; rapid data analysis and model deployment; and benefits of on-line data analysis in various real-life scenarios.

## 1.2 Sport Science and Tennis Backgrounds

At elite level in sport competitions, there is often very little difference between winning and losing. In baseball for example, a left-handed batsman can make better use of his energy from swing momentum to support transition to sprint start towards the base compared to a right-hand batsman. In golf, for right-handed players, there is a natural tendency to hit the ball with a driver producing ball trajectories that result in 'outside-in' impact (e.g. slice to the right) [6, 7]. For example, various long holes ('par five') shaped to bend to the right ('dog leg' right) would favour right-handed players who may also control better ball trajectories that curve to the right than those to the left while the opposite is true for left-handers.

Unlike golf courses, tennis courts are symmetrical, and unlike in baseball, where players run between the bases in one direction, in tennis the serve direction (from the left to the right baseline and vice-versa) is altered in each point of a game and players swap sides (i.e. 'change ends') of the tennis court on odd game scores or every six points in a tie-break game. Similarly to other studies of handedness in sport [8-10], the motivation for this study is to analyse handedness prevalence, based on available official online data. Therefore the aim of the study is to investigate if there is any higher-than-average prevalence or overrepresentation of left-handed players in tennis among elite male tennis professionals and to provide plausible explanations based on competitive game observations, empirical knowledge and hard-to-quantify descriptive common sense coaching rules.

## 2   Experimental Setup: On-line Data Collection, Tools and Procedures Followed

The data set was acquired from the official Association of Tennis Players (ATP) web site (www.atpworldtour.com, accessed 22 Aug. 2016). The data set size with top 100 tennis player rankings covers the period from 23 August 1973 – 22 August 2016 (164,217 records). The data set size was considered of sufficient size for the purpose of this study. Automated data web scraping was implemented in Python 3.x using web scraping libraries (`urllib` and `lxml`). The initial visual inspection for data elements required for web scraping code development was achieved using advanced feature shown as pop-up menu of Google Chrome web browser (**Fig. 2**). Data elements included URL and GET request method analysis for rendering web pages and table elements expressed as *XPath* – text string that is used in the Python web scraping code (**Fig. 1**).

```
plrRanking = tree.xpath ('//*[@id="rankingDetailAjaxContainer"]/table/tbody/tr/td[1]/text()')
plrMove = tree.xpath ('//*[@id="rankingDetailAjaxContainer"]/table/tbody/tr/td[2]/div[2]/text()')
plrMoveDir = tree.xpath ('//*[@id="rankingDetailAjaxContainer"]/table/tbody/tr/td[2]/div[1]/@class')
plrCountry = tree.xpath ('//*[@id="rankingDetailAjaxContainer"]/table/tbody/tr/td[3]/div/div/img/@alt')
plrPlayer = tree.xpath ('//*[@id="rankingDetailAjaxContainer"]/table/tbody/tr/td[4]/a/text()')
plrAge = tree.xpath ('//*[@id="rankingDetailAjaxContainer"]/table/tbody/tr/td[5]/text()')
plrPoints = tree.xpath ('//*[@id="rankingDetailAjaxContainer"]/table/tbody/tr/td[6]/a/text()')
plrTournPlayed = tree.xpath ('//*[@id="rankingDetailAjaxContainer"]/table/tbody/tr/td[7]/a/text()')
plrPointsDropping = tree.xpath ('//*[@id="rankingDetailAjaxContainer"]/table/tbody/tr/td[8]/text()')
plrNextBest = tree.xpath ('//*[@id="rankingDetailAjaxContainer"]/table/tbody/tr/td[9]/text()')
```

**Fig. 1.** A snippet of Python code linked to data retrieval from top 100 ATP ranking table.

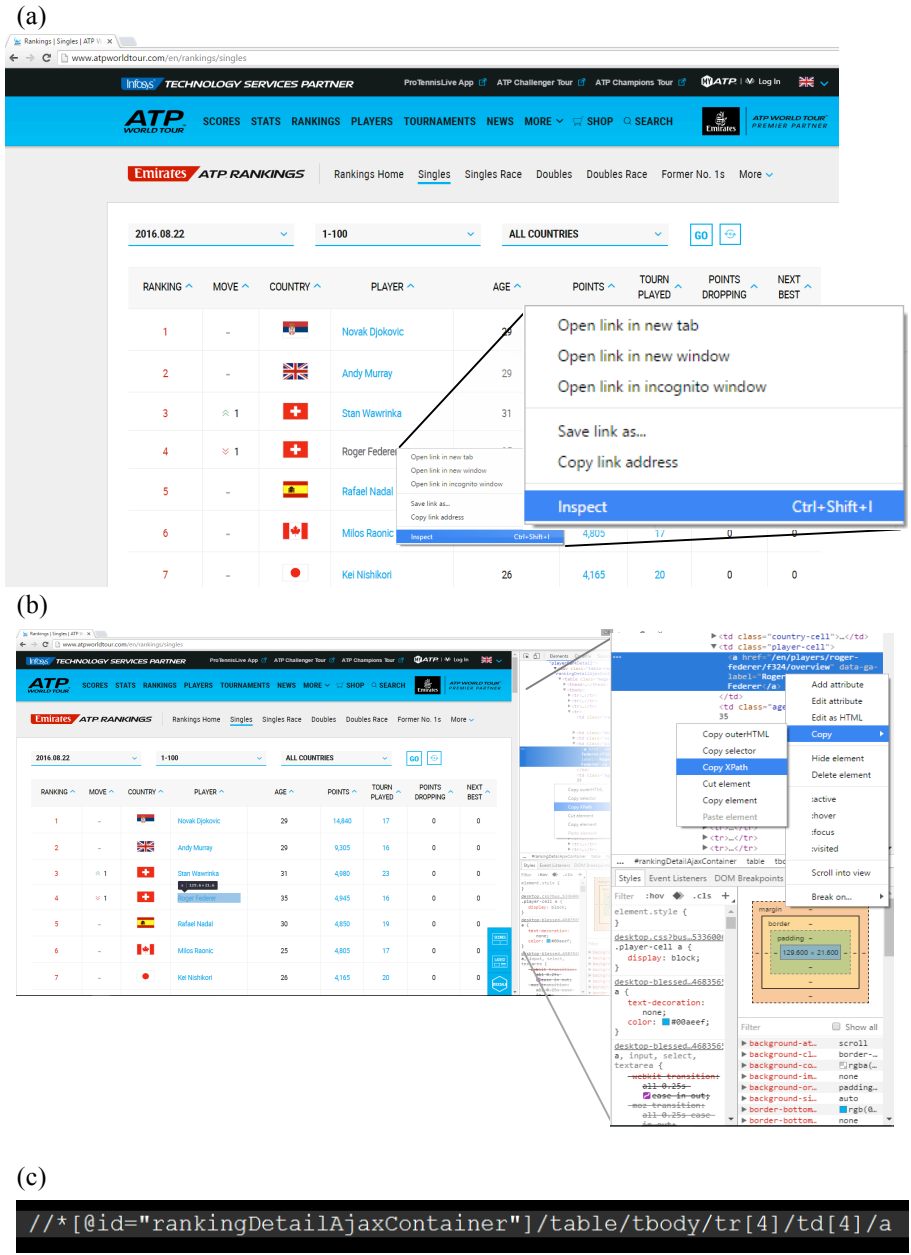

**Fig. 2.** Web page analysis of elements required for web scraping. Distinct steps cover: (a) selecting and choosing *Inspect* elements containing web page data of interest, (b) obtaining *XPath* code that is associated with the selected element and (c) copying and pasting a string of characters constituting obtained *XPath* that is associated with a selected element.

The snippet of Python code containing XPath strings (**Fig. 1**) is produced to allow data retrieval from HTML tables from the official ATP web site (**Fig. 2**).

Given that the ranking table summary does not show players' data profiles, the follow-up development stages included incremental data scraping code development, testing and further user interaction analysis aimed to find data linked to players' profiles. The development of the experimental Python code included two distinct phases: (1) retrieval and table production of top 100 ranking players since 1973, and (2) integration of players' profile data from related links. Given dynamic web sites nature (it is common for web sites to change their structure) and within the scope of this study, the design decision was that all data are collected in one table (as a non-relational database or NoSQL approach that contains data redundancy as repeated values).

The follow-up stage of research and development included decisions about data organisation, strategies to eliminate redundancy (to some degree), conversion of text into quantitative data/features, and finalising the code for importing data suitable for the scope and analysis of the study.

Given the decision regarding a NoSQL approach, the produced NoSQL table containing redundant text data was exported in various data formats (Text/CSV format, Excel *.XLS, and SQL Lite *.db) for temporary, personal and academic purposes to enable further data viewing, management, conversions and feasibility investigation for the follow-up studies using multiple environments. While preparing data for analysis, preliminary error checking revealed incomplete records and other minor errors that were considered as not significant for the scope of the study. After addressing issues such as incomplete data and outliers (e.g. two ambidextrous players' records), the final stage involved summarising data records on an annual basis.

In summary, the design framework could be described as the sequence of procedures followed for the scope of this study, covering the following distinct phases: data import, data organisation, text data conversion/quantification, and final data analysis.

## 3    Data Analysis and Results

The retrieved data set included 164,216 records covering the period from 23-08-1973 until 22-08-2016. Since some of the early records from the ATP web site included only the top 50 rather than 100 top players, after removing these incomplete early records, the data set included 156,496 records from January 1985.

**Table 1** provides a summary of the observed issues and processing decisions taken into account in support of the study.

Table 1. Findings from preliminary data set analysis and issues taken into account

| Finding importance | Description | Consequence and data processing decisions |
| --- | --- | --- |
| Low | Players may share ranking | Do not limit to 100 records/(ranking period), but to top 100 rankings. Calculate annual presence of top 100 player handedness |
| Low | Weekly and biweekly ranking Missing handedness data | Produce annual summary for left- and right-handed top players. |
| Moderate | Players categorised as: 'Ambidextrous' | Two players (Roberto Arguello and Diego Perez) were labelled as 'Ambidextrous', but no information about serving hand, switching hands or two-hand forehand preferences. Removed from analysis. Note: Fabrice Santoro records (showing: 'Right-handed, Two-Handed Backhand') were included in the analysis inspite of his unique playing style. |
| High | Missing data including availability limited to only top 50. | Records before 1985, were removed from the analysis. |

To find the annual percentage of left-handed players (**Fig. 3**), all annual top 100 rankings were summarised into the table with the following columns: year, number of records, number of left-handed players, number of right-handed players, and calculated percentage of left-handed players.

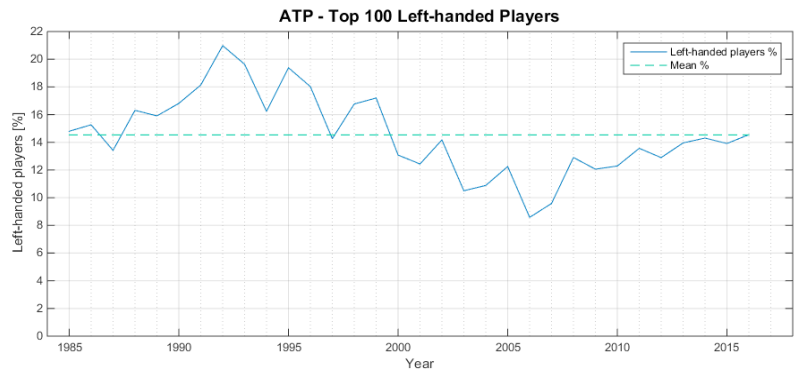

**Fig. 3.** Percentage of left-handed male tennis players in ATP top 100 ranking in the period 1985 – 2016.

While it is estimated that 10% of the world's population is left-handed, **Fig. 3** and **Table 2** show that the representation of left-handed players in elite male tennis is close to 15%.

**Table 2.** Summary of left-handed male tennis players in ATP top 100 ranking in the annual period from 1985 – 2016

| | Representation of left-handed players in top 100 |
|---|---|
| Mean | **14.53%** |
| Minimum | 8.58% |
| Maximum | 20.98% |
| Variance (Max – Min) | 12.40% |
| Standard Deviation | 2.91 |

Note: The table data were rounded to two decimal places for reader's convenience.

A plausible explanation for the trend of falling representation of left-handedness in elite tennis player rankings could be that with advancements in sport science and coaching in recent years, top tennis players may have strategically adjusted their game to better counter the playing strategies of the left-handed players. Another explanation may be related to an unknown number of surgical interventions and personal decisions which may have impacted some of the elite left-handed players' ability to compete.

## 4       Discussion

The expected trend over time is for an increase in the number of left-handed tennis players in general and can be explained with the occurrence of sociological changes (e.g. society not forcing left-handed and ambidextrous children to develop right-hand dominance), the evolving nature of tennis biomechanics and sport equipment advancements. For example, today it is common that more swings are executed with diverse stances rather than from square stance allowing faster response to aggressive attacks. On the other hand, the diversity of swing execution styles and grip styles (ways of holding the racquet for a particular swing) and racquet and string technology, has resulted in faster serving and imparting more energy to the ball that could be described as 'faster balls' and 'higher ball rotations' [11].

For enthusiasts who are new to tennis it is interesting to note that the major sport technology brands are still producing only racquets with a right-handed grip handle. Regardless of the fact that racquet frames are non-lateralised, left-hand players commonly choose to rewrap their racquets in such way so that the grip tape follows the fingers of the playing hands. Grip lateralisation is considered a minor issue for left-handed players and can be achieved by wrapping an *overgrip* tape over the original handle and/or by reengineering the racquet handle by placing the *replacement grip* tape (that is thicker than the overgrip) in opposite wrapping direction than for right-handed grip. At present, the major grip tape manufacturers (with the exception of *Tourna* grip) produce their grip tapes pre-cut for right-handed players and with a single side to be in contact with the skin. By cutting the end of the grip tape for left-handed players, it also means that a player is effectively losing a small part of the grip tape length available. Similar to preferences for holding a racquet, re-gripping preferences are also individual as well as the shape of the racquet handle.

The increase in prevalence of right-handed player evident in this study is limited to the male population. Taking into account factors such as common knowledge of diversity of male vs. female players, un-returned serve percentage based on gender, and to some extent diverse on-line data sources, the authors' view is that all these factors would warrant another study that may lead to new discoveries.

One of the limitations of the web scraping approach is that the web scraping code typically needs modifications when a web site containing data changes its' internal structure. In this study, the experimental design strategy was oriented towards obtaining data to support a specific research question rather than copying and producing a local version of a normalised SQL database which would allow long-term further research without additional data retrieval. The disadvantage of a short and simple web scraping code is that the obtained data contains duplicate and redundant elements.

Regarding observations of tennis matches and common-sense descriptive rules, **Table 3** provides a summary of plausible key points and rationale regarding the benefits of left-handedness in tennis. This summary can assist in informing coaching and parents' decisions regarding the motivation whether to encourage the starting left-handed playing preference (when circumstances allow).

Table 3. Empirical and common-sense coaching rules from game analysis involving left-handed player

| Key advantage | Description |
| --- | --- |
| • Game-deciding points with wider serving options<br>– Serve:<br>  o reward for slice serve trajectory where ball is passing close to the lowest point of the net height<br>  o out-wide serve on right-handed backhand | Due to the Magnus effect of the ball spin, point counting and court geometry are in favour of left-handed players with more important points on the even side (**Fig. 4**, side II) i.e. defending the break point or winning the game point. In contrast, for right-handed players, the favoured side I (**Fig. 4**) creates an advantage point when both players have the same score. The out-wide serve that does not land close to the line is less penalised with the right-handed backhand than when the right-handed player serves the out-wide serve to another right-handed player. |
| – Return:<br>  o reward for left-hand forehand where it is easier to | Similar to the importance of the serving side is that right-handed players are more vulnerable on game-decision points (**Fig. 4**, side II) to |

| Key advantage | Description |
| --- | --- |
|     return a wide serve than for right-handed backhand<br>   o penalty for down-the line right-handed backhand return where the ball is passing to the highest point of the net height requiring height, direction and depth control of the ball<br>   o penalty for diagonal right-handed return where a greater margin for error also means placing the ball in the zone for left-handed opponent's forehand winner | aggressive forehand returns from left-handed opponents if an insufficiently wide serve or the second serve is within the returner's comfort zone. In contrast, right-handed backhand diagonal returns that have a greater margin of error than down-the-line shots are also more likely to land in a position where the left-handed server can execute a forehand winner. With long matches where there is little difference between the players, second serve variations on game-decision points cause elevated stress levels, particularly for right-handed servers when defending the game point (i.e. to serve to stay within the game). |
| • Top-spin and slice trajectories | The ball trajectories and attack angles produced by the left-handed minority are more difficult to interpret for the majority of right-handed players. |
| • Cross-court rally assertiveness | Developing assertiveness to switch the cross-court direction should be easier for left-handed players than for right-handed players. The minority of players have more opportunities to practice and compete against the majority. |
| • Minority practicing against majority | Left-handed game philosophy is developed naturally over the years. For right-handed players to play well against the left-handed players they need regular practice and coaching guidance aimed at neutralising left-handed opponents' game advantages. |
| • Net opportunities from cross-court and out-wide | Backhands in general have less reach than forehands, so players on the backhand side often do not cover the court well after returning a wide backhand. |
| • Doubles team play | A left-handed partner can bring versatility to the team. Right-handed players have less left-handed players to choose as their doubles partners. Left handed players being in greater demand for double partners may ended up teaming up with higher-ranked partner than themselves. |

For recreational players, in the first author's view (who is also a certified tennis coach), it is beneficial to include left-hand in occasional practise and include left-hand in warm up to promote muscle balance and proprioceptive ability.

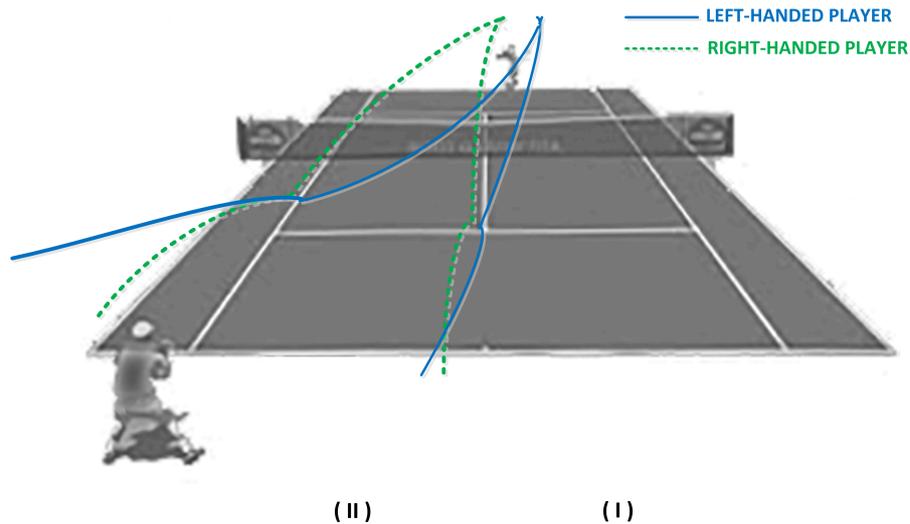

**Fig. 4.** Left- and right-handed ball trajectories and wider left-handed serving angles for even game points on side (II). For side (I) a right-handed server has wider serving angles compared to the left-handed player.

Learning to serve left-handed may be strategic advantage with some points when the opponent's stress level is higher than yours and when you or your doubles team can afford a first-serve mistake.

Right-handed players at all skill-levels should consider studying and adapting mirrored left-handed strategy as a game plan against left-handed opponents. For example, when scouting a left-handed player who is shorter than 180 cm – it is common to expect slice serve and variations in serve trajectories and general ball bounce. Visualising the ball trajectory should positively contribute to hand-eye coordination, which in turn will help adjust quicker to the left-handed opponent during the match. Consider practicing with opposite hand (i.e. left hand for right-handed players) for short time durations and learn to be patient with yourself. Practice mirrored left-handed strategies (and if it does not come naturally, consider practicing it against less-experienced left-handed players or when you have a 'nothing to lose' situation during a match).

Regarding the technical aspect of the text data mining component of the study, it is interesting to note that Python 3.x uses web scraping libraries (`urllib` and `lxml`) that can retrieve tree data structures at once. This is in contrast to Python 2.7 library (`requests`) that requires an iterative approach rather than a single command. The iterative approach, however, allows programmers to access related data within the same data retrieval loop. For most operating systems, and Spyder *integrated development environment* (IDE) installations, Python 2.7 is a default version, which represents an inconvenient incompatibility and causes installation problems if there are mutually exclusive dependencies. For example, Ubuntu 14.04.x LTS with support until 2019 does not allow coinciding installations of Spyder 3x and Spyder 2x IDEs. For users who disagree with Ubuntu 16.x LTS privacy end user agreement (EULA), it is advised to consider using multiple virtual machines or installing e.g. the latest 64-bit distribution of Linux Mint 17.3 LTS (supported until April 2019) that passed our tests regarding dual and coinciding installation of Spyder 3.x and 2.x IDEs and Matlab R2016a (used for analysis and to produce **Fig. 3**).

## 5   Conclusions, Recommendations and Future Work

Compared to the estimates of about 10% of people being left-handed in the general male population, data analysis from the official ATP web site over the past three decades (1985-2016) demonstrate that left-handed elite tennis players account for about 15% (varying from 8.6–21%) of the top 100 players. The insights and data analysis can inform parents, coaches, strategic game concepts, enhance media coverage/analytics, left-handed facts and statistics, and inform tennis equipment manufacturing. The demonstrated experimental setup and multi-disciplinary insights based on web data retrieval show the benefits of expanding text mining into sport science, coaching, tennis strategy, media coverage and sport technology.

**Acknowledgements**. The authors wish to express their appreciation to ATP developers (www.atpworldtour.com) for providing error-free public data viewing through the history of tennis, the designer teams of Python web scraping/parsing library (lxml and urllib) and Spyder IDE (https://pythonhosted.org/spyder/) utilised in this study. This study is limited to academic data analysis, data summarisation and interpretation, and it does not facilitate any data retrieval system to third parties. The authors also wish to express their sincere appreciation to Assoc. Prof. Russel Pears for his valuable comments and encouragement.